\documentclass{scrartcl}
\usepackage{graphicx}
\usepackage{cite}
\usepackage{authblk}

\usepackage{hyperref}
\usepackage[utf8]{inputenc}

\newcommand{\eqref}[1]{(\ref{#1})}

\date{}

\begin{document}
\title{Maxwell demons in phase space}
\author{Juan M.R. Parrondo and  L\'eo Granger}
\affil{Departamento de F\'\i sica At\'omica, Molecular y Nuclear and GISC, Universidad Complutense de Madrid, 28040,  Madrid, Spain}
\maketitle
\begin{abstract}
	Although there is not a  complete ``proof'' of the second
	law of thermodynamics based on microscopic dynamics, two
	properties of Hamiltonian systems have been used to prove the
	impossibility of work extraction from a single thermal
	reservoir: Liouville's theorem and the adiabatic invariance of
	the volume enclosed by an energy shell. In this paper we
	analyze these two properties in the Szilard engine and other
	systems related with the Maxwell demon. In particular, we recall
	that the enclosed volume is no longer an adiabatic invariant
	in non ergodic systems and explore the consequences of this on
	the second law.
\end{abstract} 

\section{Introduction}

Since the second half of the last century it is accepted that the
second law of thermodynamics --- and consequently the Carnot
principle --- has a probabilistic nature and is only valid in average.
A Brownian particle in a field, for instance, can gain potential
energy from the surrounding fluid converting heat into work for short
periods of time. Already in 1860 Maxwell realized that these
fluctuations can be used to beat the second law and illustrated this
idea with a celebrated gedanken experiment: the {\em Maxwell demon}
\cite{Leff} .

The Maxwell demon was the starting point for a fruitful exploration of
the relation between entropy and information, which is rooted in the
foundations of statistical mechanics and thermodynamics. Szilard
devised a stylized version of the demon, the {\em Szilard engine}, in
which this relation could be  quantified: one bit of information (the
outcome of a yes/no experiment) reduces the entropy by $k\ln 2$ and,
for instance, allows one to extract an energy $kT\ln 2$ from a single
thermal bath in a cyclic process. 

After Maxwell and Szilard, much effort was devoted to assess the
entropic cost of measurement that could restore the validity of the
second law in the presence of feedback. However, in the 1970's,
Landauer and Bennett introduced a novel and surprising element: an
entropic cost in the erasure of the information that the demon gathers
in the measurement. Landauer showed that the erasure or, more
precisely, the isothermal restoring of a memory to a reference state,
dissipates heat, increasing the entropy of the surrounding thermal
reservoir  \cite{Landauer1961}. Bennett went further  and applied the
{\em Landauer principle} to the Szilard engine, proving that for some
setups the measurement can be realized at zero entropy cost and the
second law is restored by the cost of erasure  \cite{Bennett1982b}.

In the last years, the thermodynamics of information has experimented a
new development with the application of non-equilibrium fluctuation
theorems and stochastic thermodynamics
\cite{Sagawa2013inbook,Sekimoto,gaveau_relative_2014,parrondo_thermodynamics_2015}
and also due to the
experimental verification of the Landauer principle \cite{Berut2011}
and several experimental realizations of the Szilard engine
\cite{Toyabe2010,Koski2014b,Roldan2014}.

More recently, two atypical examples of Szilard engines were
introduced \cite{Marathe:2010fx,Vaikuntanathan:2011en}. Unlike the
classical Szilard engine, which operates with a feedback loop and in
contact with a heat bath, these systems can be cooled down during a cyclic
process {\em without feedback}.
They are called {\em microcanonical} Szilard engines, because they work
if the system is prepared in a microcanonical initial state and is
thermally isolated during the process.
Despite these differences, they  share a fundamental feature with the original
Szilard engine, namely, at some stage of the cycle they undergo a
{\em symmetry breaking}. In other words, the ergodicity of the system
is lost implying that the phase space of
the system gets split into disconnected regions. 
The fact that symmetry breaking is the key ingredient of the thermal,
feedback driven Szilard engine was already pointed out in
\cite{parrondo2001szilard} and explored further in \cite{Roldan2014}.

From the phase space point of view, two fundamental properties of
Hamiltonian systems forbid work extraction from a single heat source
during a cyclic process: {\it i)} the conservation of phase space
volume, known as {\em Liouville's theorem}, and {\it ii)} the
adiabatic invariance of the volume enclosed by an energy shell.
Therefore, during a cycle of a Szilard engine,
at least one of these
two properties must be broken.

In this paper,
we attempt an analysis of generic Maxwell demons,
thermal and microcanonical, from a phase space perspective. Our aim is
to characterize the behavior of trajectory bundles in phase space
under feedback and non-feedback Hamiltonian evolutions, in which
energy shells split into disconnected regions at some stage of the
process.
We analyze the consequences of this splitting on
the aforementioned properties {\it i)} and {\it ii)}.
This novel approach is particularly suited for
the present special issue, since Professor Jacques Yvon contributed
greatly to our understanding of the role of phase space dynamics in
statistical mechanics through the celebrated BBGKY hierarchy.

The paper is organized as follows. In section \ref{sec:dyn} we review
the basic properties of Hamiltonian dynamics in phase space with
special attention to Liouville's theorem and the adiabatic invariance
of the volume enclosed by an energy shell. In section
\ref{sec:liouville} we analyze the behavior of trajectories in phase
space for the original Szilard engine and discuss how feedback allows
one to beat Liouville's theorem and effectively contract a region in
phase space. In section \ref{sec:helm} we study the consequences of
the failure of the adiabatic invariance of the enclosed volume  in non
ergodic evolution and the restrictions that Liouville's theorem
imposes to cyclic non feedback protocols in phase space. Finally, in
section \ref{sec:conc} we present our conclusions and discuss the
subtle problem of the definition of entropy for single systems,
instead of ensembles, when ergodicity is broken.

\section{Basic phase space dynamics}\label{sec:dyn}

The solution of the Hamilton equations in the phase space $\Gamma$
defines an evolution operator $(q(t'),p(t'))=U_{t',t}(q(t),p(t))$.
This evolution operator satisfies two restrictions which are characteristic of Hamiltonian dynamics: the Liouville's theorem
and the adiabatic invariance of the volume enclosed by an energy shell. Liouville's theorem states that
the volume ${\rm vol}(A)$ of a set of micro states $A\subset \Gamma$ is
invariant, i.e.,  ${\rm vol}[U_{t',t}(A)]={\rm vol}(A)$. 

The adiabatic
invariance of the enclosed volume is more restrictive and only applies to ergodic systems undergoing quasistatic processes. In these processes, the system evolves under a time-dependent Hamiltonian $H(q,p;\lambda_t)$, where $\lambda_t$ is one or several external parameters that change infinitely slowly in time $t=[0,\tau]$, that is $\dot \lambda_t\to 0$ and $\tau\to \infty$. For the following discussion it is useful to distinguish between an {\em energy shell} of energy $E$
\begin{equation}
\gamma_E\equiv\{ (q,p):H(q,p;\lambda)=E\}
\end{equation}
and an {\em energy layer} of finite width $\Delta E$:
\begin{equation}
\Gamma(E;\Delta E)\equiv \{ (q,p):E\leq H(q,p;\lambda)<E+\Delta E\}.
\end{equation}

In a quasistatic process where the system remains ergodic on every energy shell,  $\gamma_E$
is entirely mapped into another shell $\gamma_{E+W(E)}$, $W(E)$ being the
work done on the system along the process which generically depends on the
initial energy. The work $W(E)$  is deterministic because the
system explores the whole energy shell at each stage of the
process, due to quasistaticity and ergodicity together. Hence, a
point $(q,p)$ in $\gamma_E$ ends the process with an energy $E+W(E)$
lying in $\gamma_{E+W(E)}$. Reversing the momenta $(q,p)\to (q,-p)$ and
the protocol followed by the external parameter, one concludes that, if
$H(q,p;\lambda)=H(q,-p;\lambda)$, then every point in $\gamma_{E+W(E)}$ is mapped by this
reverse process into a point in $\gamma_{E}$. Therefore, we conclude
that $U_{\tau,0}[\gamma_{E}]=\gamma_{E+W(E)}$. Let us define the volume of
the subset enclosed by  $\gamma_E$  as
\begin{equation}
\phi_\lambda(E)\equiv\int_\Gamma dqdp\,\Theta(E-H(q,p;\lambda)).
\end{equation}
Ergodicity also implies that the energy layers do not cross each other along the process, i.e., that  $E'+W(E')>E+W(E)$ for all $E'>E$. In other words, the final energy $E+W(E)$ is a strictly monotonic increasing function of the initial energy $E$. Then  Liouville's theorem applied to the enclosed volume implies
\begin{equation}\label{adiab}
\phi_{\lambda_0}(E)=\phi_{\lambda_\tau}(E+W(E)).
\end{equation}
In particular, for a cyclic process, $\lambda_0=\lambda_\tau$, $W=0$, since $\phi_\lambda(E)$ is a one-to-one function of $E$.

For  non-equasistatic processes, the work is no longer deterministic
and an energy shell is mapped onto  a more intricate set that
intersects several shells ranging from $E+W_{\rm min}$ to $E+W_{\rm max}$, $W_{\rm min}$ and $W_{\rm max}$ being the minimum and maximum value of the work.  The usual assumption in statistical mechanics is that the intricateness of this final set cannot be described in terms of meso- or macroscopic states: any coarse graining in phase space effectively smears the set to cover the whole energy layer of a width $W_{\rm max}-W_{\rm min}$.  One can conceive  other mechanisms of information loss with the same result: for instance, if, after completing the process, we let the system evolve for a random time with the time independent and ergodic Hamiltonian $H(q,p;\lambda_\tau)$, then the final set of an ensemble of systems starting in $\gamma_E$ will be a layer of finite width $W_{\rm max}-W_{\rm min}$. In any case, the effective volume of a set increases in a non-quasistatic processes.

Summarizing, quasistatic processes map energy layers into energy layers with the
same volume, whereas non-quasistatic processes map  energy layers into
intricate sets that any information loss or relaxation mechanism
effectively convert into layers with a larger volume than the original
one. 
The apparent violation of the second law by a Maxwell demon should be related with either Louville's theorem  or the adiabatic invariance of the enclosed volume $\phi(E)$  at the level of phase space dynamics. In the next sections we analyze how several Maxwell demons proposed in the literature operate in the phase space and compromise the validity of Liouville's theorem and the adiabatic invariance of the enclosed volume.

\section{The Szilard engine in phase space}\label{sec:liouville}

\begin{figure}
\[
\includegraphics[width=0.7\textwidth]{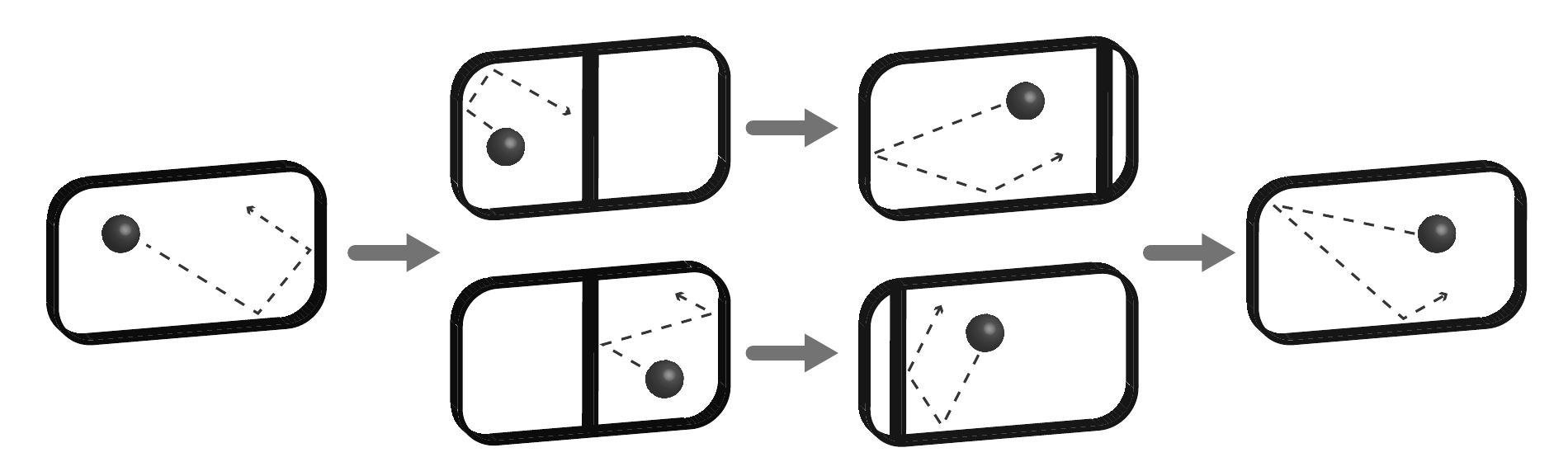}\]
\caption{The Szilard engine. A demon insterts a piston in the middle of a container with a single-molecule gas surrounded by a thermal bath at temperature $T$. The demon measures in which half the particle lies and carries out a reversible expansion extracting a work $kT\ln 2$ from the thermal bath in a cycle. Notice that the protocol followed by the demon depends on the outcome of the measurement.}
\label{figszilard}
\end{figure}

The Szilard engine consists of a single particle gas at temperature $T$: a particle in a container of volume $V_0$ whose velocity after a collision with the walls is randomly set according to a Maxwell distribution at temperature $T$. As sketched in Fig.~\ref{figszilard}, an external agent or demon inserts a piston in the middle of the container, measures in which half the particle lies and performs a reversible expansion from a volume $V_0/2$ to $V_0$, extracting an amount of work
\begin{equation}\label{w1}
W_{\rm extr}=\int_{V_0/2}^{V_0} PdV=kT\ln 2
\end{equation}
where we have used the state equation $PV=kT$, $P$ being the average pressure exerted by the particle and $k$ the Boltzmann constant.  The measurement is necessary to implement the reversible expansion, since the external agent must exert a force identical to the pressure of the single-molecule gas \cite{Leff}. It is important to notice  that the Szilard engine can be realized with any system that undergoes a splitting of its phase space~\cite{parrondo2001szilard} ---like a colloidal particle in a double well potential \cite{Kawai:2007kc,Roldan2014} or electrons in quantum dots~\cite{Koski2014b}---, using an appropriate feedback protocol.

\begin{figure}
\vspace{0.4cm}
\centering
\includegraphics[width=0.8\textwidth]{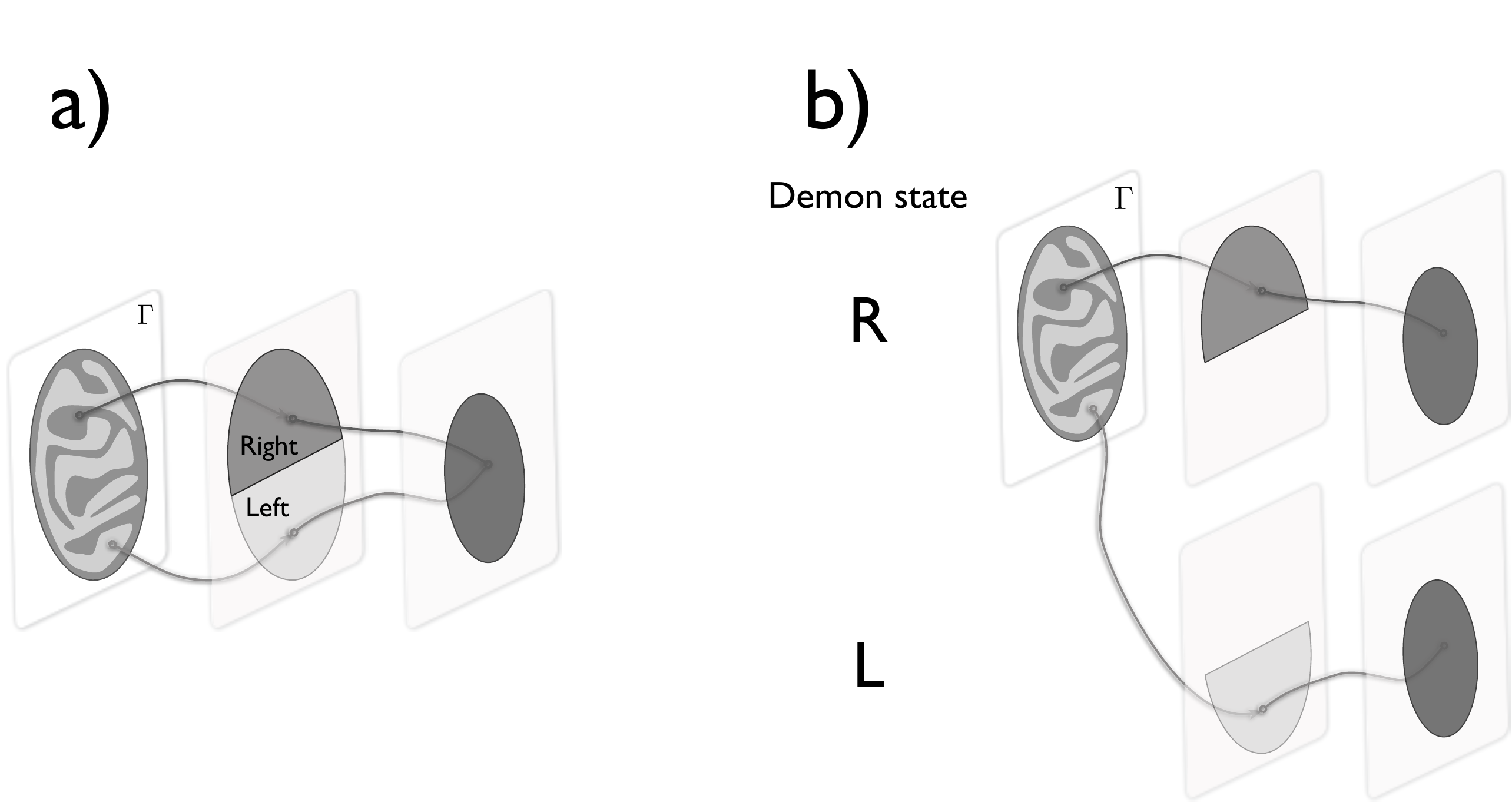} 
\caption{Trajectories in the phase space $\Gamma$ of the single molecule gas plus the thermal reservoir in the Szilard engine. a) The feedback in the cyclic protocol implies that two trajectories with different initial conditions trajectories cross at the end of the process decreasing the volume in phase space by a factor 2. b) The state of the demon is included in an extended phase space $\Gamma_{\rm tot}=\Gamma\times \{L,R\}$. The demon is initially in state $R$ and the trajectories are identical as in the previous case a), but now the demon must be restored to state $R$ to complete the cycle.}
\label{fig:trajectories}
\end{figure}

At the level of the phase space, the feedback control allows trajectories to cross each other, inducing a reduction of phase space volume. This is sketched in Fig.~\ref{fig:trajectories} a) where $\Gamma$ represents the phase space of the single molecule gas {\em and} the thermal reservoir.  Any final microstate $(q_\tau,p_\tau)$ is reached by two different trajectories starting at $U_L^{-1}(q_\tau,p_\tau)$ and $U_R^{-1}(q_\tau,p_\tau)$, where $U_{L,R}$ is the evolution operator $U_{\tau,0}$ corresponding to the protocol implemented when the particle lies in the left and right half of the container, respectively. According to Liouville's theorem, this means that a subset in the phase space of volume $\phi_{\rm init}$ shrinks to  a final volume $\phi_{\rm fin}=\phi_{\rm init}/2$. If the final energy $E+W(E)$ is a monotonous function of the initial one and the process is quasistatic, then the region enclosed by the energy shell $\gamma_{E}$, with volume $\phi(E)$ is mapped onto the region enclosed by a shell  $\gamma_{E+W}$  obeying 
\begin{equation}
\phi(E+W)=\frac{\phi(E)}{2}
\end{equation}
Taking logarithms and expanding to first order in $W$:
\begin{equation}\label{w2}
\frac{W}{kT}=-\ln 2
\end{equation}
since $\frac{d\ln\phi(E)}{dE}=\frac{1}{kT}$, which is the result that we already obtained applying the state equation of ideal gases to the single molecule gas (cfr. \eqref{w1} and \eqref{w2}, with $W_{\rm extr}=-W$). 

Liouville's theorem is recovered if we take into account the physical nature of the demon and 
analyze trajectories in an extended phase space $\Gamma_{\rm tot}=\Gamma \times \{ L,R\}$, which 
includes the state of the demon, $L$ or $R$, depending on the outcome of the measurement. These are not 
necessarily microscopic states. They can be meso- or macroscopic states covering equal volumes in 
the phase space of the microstates of the demon (see \cite{Sagawa2009} for an example with a non 
symmetric demon). In Fig.~\ref{fig:trajectories} b) we show trajectories in which the demon starts in a definite 
state $R$. The trajectories in $\Gamma$ are identical as the ones discussed above, but now the 
final set covers, in the extended phase space $\Gamma_{\rm tot}$, the same volume as the initial one. 
The total system does not complete a cycle, since the demon, originally in state $R$, ends either in 
$L$ or $R$. To complete the cycle one has to reset the demon to the original state $R$ which implies 
a reduction of the phase space volume that has to be compensated by an increase in the phase space 
volume of the thermal bath. This increase is the heat dissipation associated to erasure, as dictated by 
Landauer's principle.

\section{Beating  the adiabatic invariance of $\phi(E)$}\label{sec:helm}

In the derivation of equation~\eqref{adiab} and the discussion of the previous section, we have shown that the impossibility of extracting work in a cycle relies on two basic facts: Liouville's theorem and the monotonicity of the final energy $E+W(E)$ as a function of $E$. This monotonicity is warranted if the energy shells are compact connected sets along the whole process, i.e., if the shells consist of a single closed hypersurface in $\Gamma$. In that case, the shells cannot cross each other.  On the other hand, the Szilard engine is an example of a quasistatic processes where energy shells split into several disconnected components. We will show that in those cases Liouville's theorem is still valid but the enclosed volume $\phi$ could be no longer invariant.

Consider a processes where ergodicity is broken at some stage and energy shells $\gamma_E$ split into disconnected regions $i=1,2,\dots$ (the number of regions could depend on the initial energy $E$). Even if the process is quasistatic, the work is no longer deterministic but can take different values $W_i(E)$ with probability $p_i(E)$, which is the probability that the system occupies region $i$ in the non-ergodic stages of the process.

We are interested on how the enclosed volume $\phi_{\lambda_0}(E)$ is transformed under this non ergodic evolution. Let  ${\cal U}_i(\phi_{\lambda_0}(E)) \equiv\phi_{\lambda_\tau}(E+W_i(E))$ be the transformation of the enclosed volumes $\phi$. Consider  an initial layer $\Gamma(E,\Delta E)$ and let $\Delta\phi_0=\phi_{\lambda_0} (E+\Delta E)-\phi_{\lambda_0}(E)$ be its volume. The volume of the set of initial microstates $(q,p)$ in the layer that cross region $i$ along the process is $p_i\Delta \phi_0$. Now consider the time-reversed protocol with initial condition in the final layer $\Gamma(E+W_i(E),\Delta_i E')$. Let  $\tilde p_i(E)$ the fraction of those states that go back to the initial layer $\Gamma(E,\Delta E)$.  Liouville's theorem implies:
\begin{equation}
p_i\Delta\phi_0=\tilde p_i\left[ {\cal U}_i(\phi_{\lambda_0}(E)+\Delta\phi_0)-{\cal U}_i(\phi_{\lambda_0}(E))\right].
\end{equation}
The limit $\Delta\phi_0\to 0$  immediately yields
\begin{equation}\label{slopes}
{\cal U}'_i(\phi_{\lambda_0}(E))=\frac{p_i(E)}{\tilde p_i(E)}
\end{equation}
which is  a condition that any non-feedback evolution must fulfill. Other than that, the transformation of $\phi_{\lambda_0}(E)$ can be in principle arbitrary.

\begin{figure}
\vspace{0.4cm}
\centering
\includegraphics[width=0.95\textwidth]{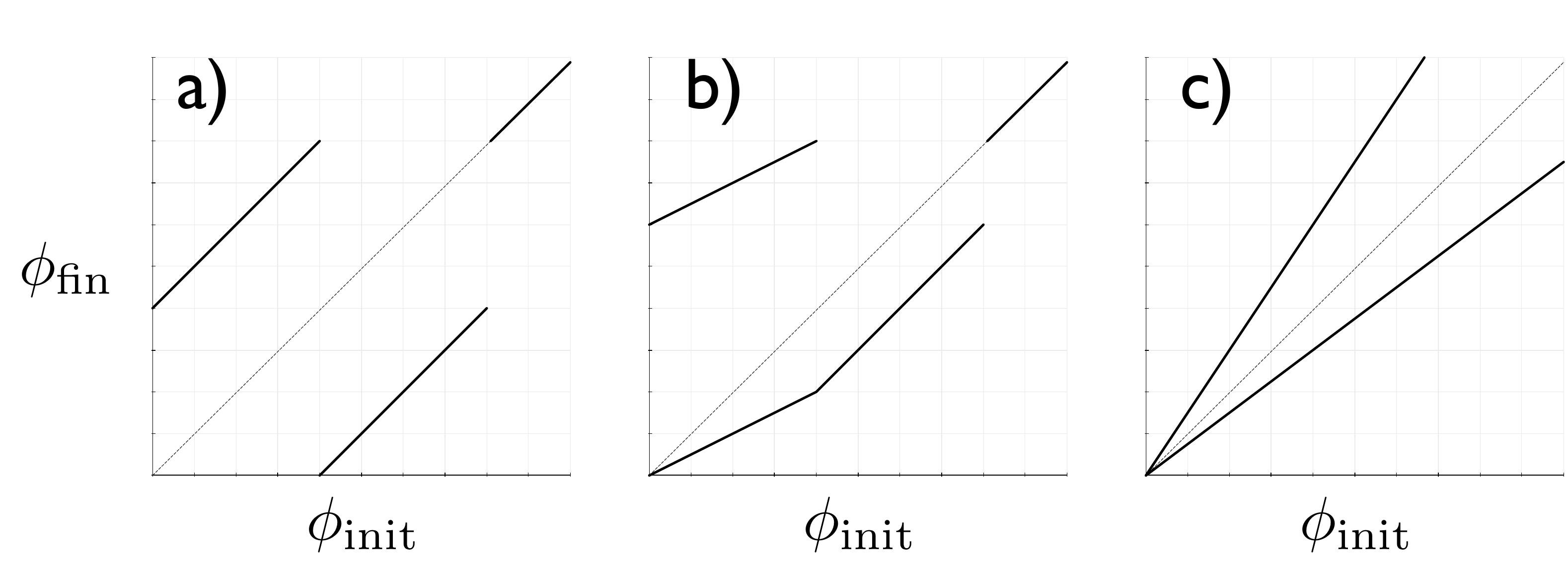} 
\caption{Three examples of the transformation of the volume enclosed by energy shells. The initial volume $\phi_{\rm init}\equiv \phi_{\lambda_0}(E)$ is mapped into $\phi_{\rm fin}\equiv \phi_{\lambda_\tau}(E+W_i(E))$:  a) corresponds to the microcanonical Szilard engine introduced by Vaikuntanathan and Jarzynski \cite{Vaikuntanathan:2011en} with $p_i=\tilde p_i=1$ (see Fig.~\ref{fig.vaikuJar}); b) corresponds to the microcanonical Szilard engine introduced by Marathe and Parrondo \cite{Marathe:2010fx} with $p_i=1/2$ ($i=L,R$) and $\tilde p_i=1$ (see Fig.~\ref{f.microSz}); c) corresponds to a Szilard engine in contact with a thermal bath at temperature $T$ with $p_L=p_R=1/2$  and $\tilde p_L=2/3$, $\tilde p_R=1/3$ (see Fig.~\ref{fig.ebproc}). It is easy to check that in all cases the slopes verify Eq.~\eqref{slopes}.}
\label{fig:actions}
\end{figure}

\begin{figure}
	\includegraphics[scale=0.85]{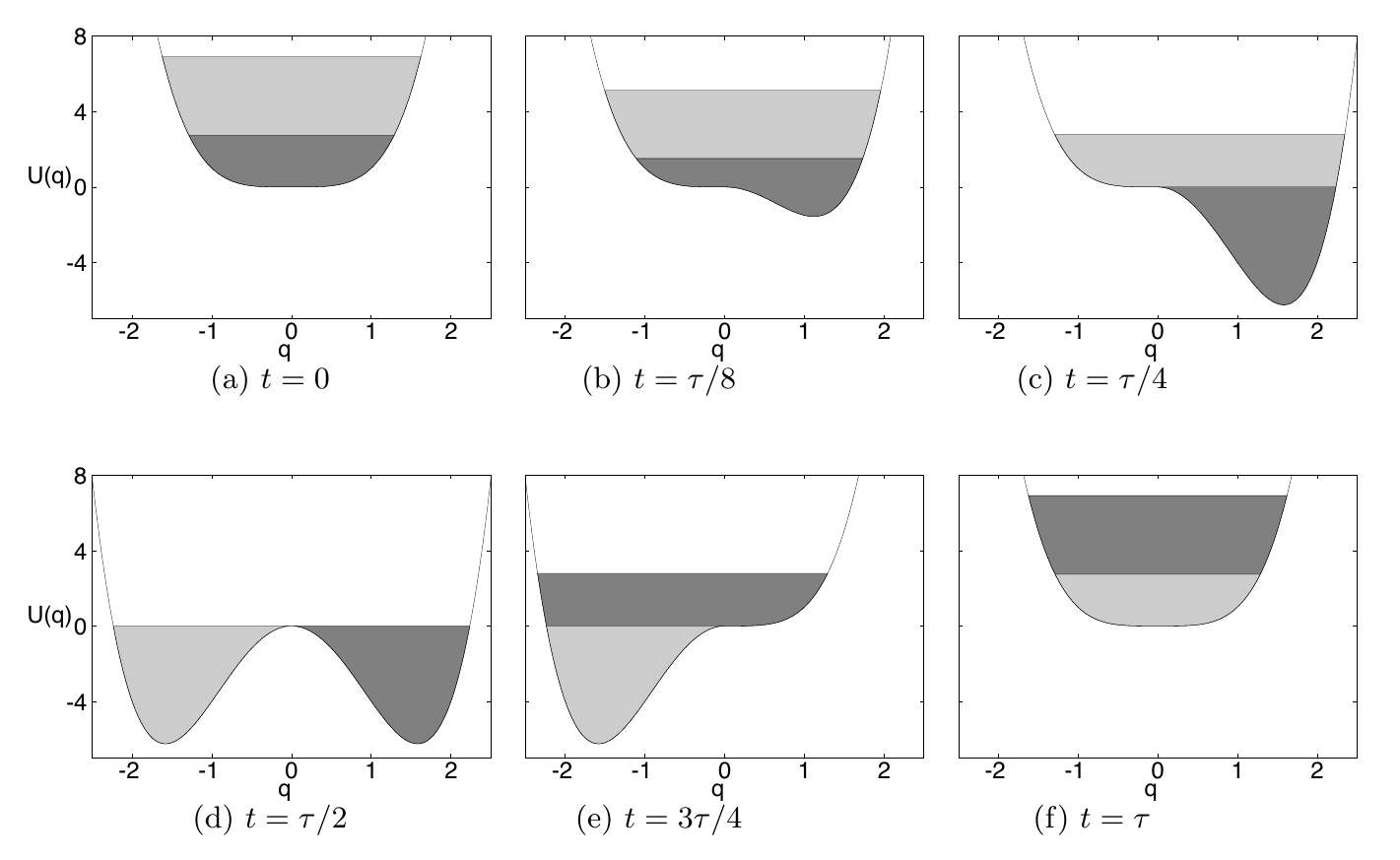}
	\caption{The microcanonical Szilard engine by Vaikuntanathan and
		Jarzinski in \cite{Vaikuntanathan:2011en}. The model
	consists of a classical particle subjected to a one
dimensional quartic potential $U(q)$. The panels also show the evolution of regions in phase space formed by points with a given coordinate $q$ (horizontal axis) and total energy (vertical axis).
Panel (a): Initially, the potential is unimodal
and the system is ergodic. The cycle proceeds as follows. Panels (b)
and (c): a well is slowly created on the right part of the potential. Panel (d):
a second well is slowly created on the left part of the potential. At
that stage, the potential is bimodal and the ergodicity of the system
is broken. Panel (e): the energy of the right well is slowly increased
until the well disappears. Panel (f): The left well disappears in the same
way and the potential recovers the initial shape. The net effect of this cycle is to swap the two regions of phase
space shaded in dark and light grey on the figure.}
	\label{fig.vaikuJar}
\end{figure}

In Fig. \ref{fig:actions} we present three possible transformations ${\cal U}_i$ that correspond to physical models introduced in previous works.
Figure \ref{fig:actions} a) corresponds to the model introduced by
Vaikuntanathan and Jarzynski in \cite{Vaikuntanathan:2011en}.
In this model, an isolated one-dimensional  classical
particle is subjected to a cyclically varying quartic potential.
The protocol is sketched on Fig.~\ref{fig.vaikuJar}.
The effect of the cycle is to interchange two energy layers (depicted
in dark and light grey on Fig.~\ref{fig.vaikuJar}).
This process involves ergodicity breaking as can be seen on  panel
(d) of Fig.~\ref{fig.vaikuJar}.
The work performed during this cycle is a deterministic function of
the initial energy. Therefore $p_i = 1$.
Since the protocol is symmetric in time, we also have $\tilde{p}_i =
1$, leading to a slope equal to one on Fig.~\ref{fig:actions} a), even
though the curve is only piecewise continuous.

\begin{figure}
	\includegraphics[scale=0.7]{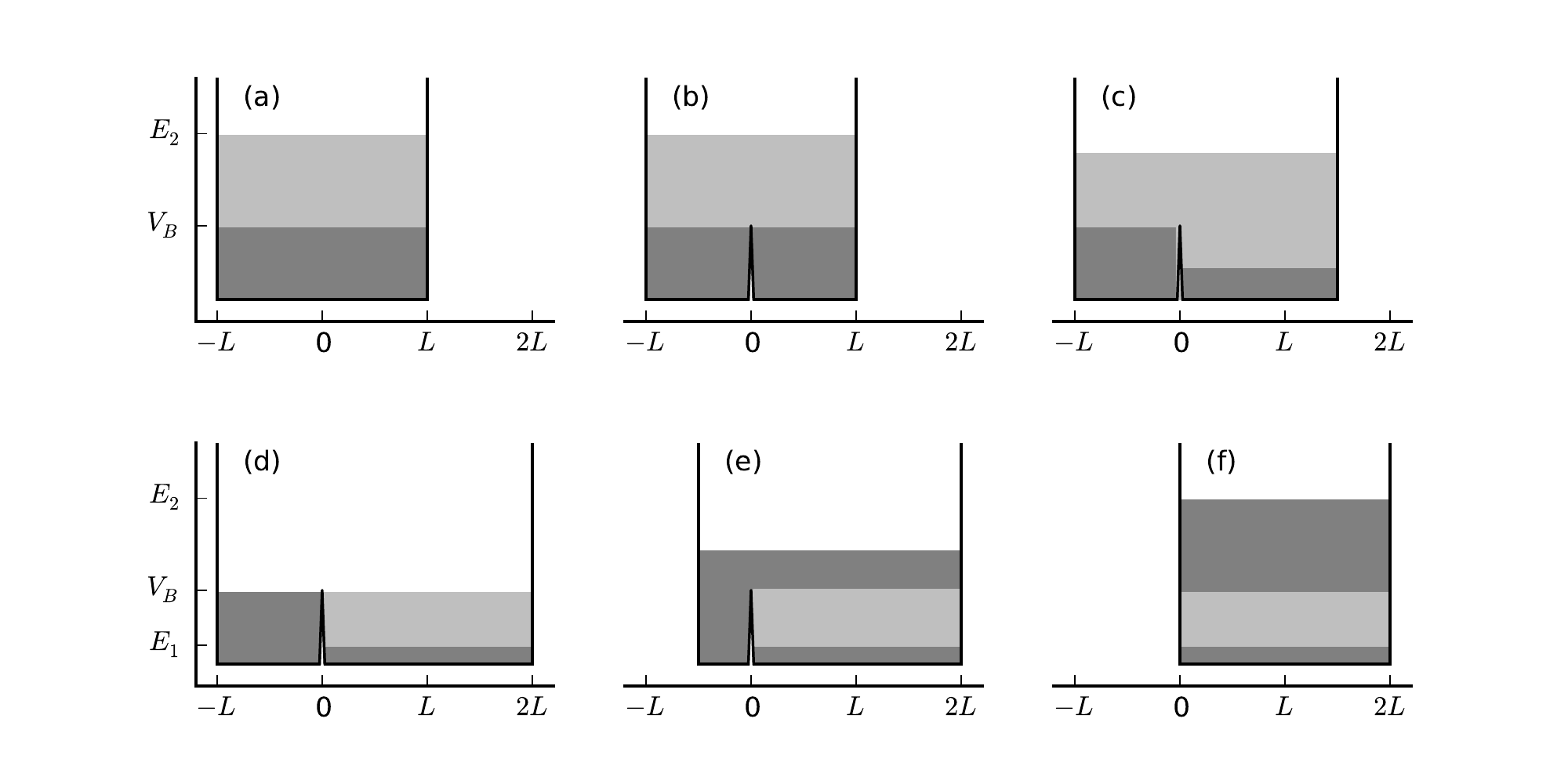}
	\caption{Microcanonical Szilard engine introduced by Marathe
		and Parrondo in \cite{Marathe:2010fx}. The system consists of a classical
particle confined in a one dimensional box and subjected to a potential that changes quasistatically. As in Fig.~\ref{fig.vaikuJar}, the panels show the potential and the evolution of regions in phase space. The cycle
	proceeds as follows. First, a
potential barrier of height $V_B$ is introduced in the middle of the
box (panel (b)). Then, the right wall is slowly moved to the right
(panels (b,c,d)). Finally, the left wall is slowly moved to the
position of the barrier (panels (d,e,f)). The effect of this cycle is to
transport the energy layers as depicted on panel (f). In particular,
the layer of energy lower than $V_B$ gets split into two disconnected
layers, the dark grey regions in panel (f).}
	\label{f.microSz}
\end{figure}

Figure \ref{fig:actions} b) corresponds to a model introduced by Marathe
and Parrondo \cite{Marathe:2010fx}.
The cycle is depicted on Fig.~\ref{f.microSz}. 
This model consists of a classical particle in a one dimensional box
of size $2L$. The cycle proceeds as follows. A potential barrier of
height $V_B$ is inserted in the middle of the box. Then, the right wall
is moved quasistatically slowly to the right. Finally, the left wall
of the box is moved to the position of the barrier.  When the barrier
is inserted, ergodicity is broken.  If the energy of the particle is
smaller than the height of the barrier, then the work performed
depends on whether the particle is trapped on the left side or on the
right side of the barrier. Each of these cases happen with equal
probability $p_{\rm L} = p_{\rm R} = 1/2$.  In fact, any shell of
energy lower than $V_B$ gets split into two disconnected shells in the dark grey area in Fig.~\ref{f.microSz} f).
However, in the reverse process the work
performed is deterministic.  Hence, $\tilde{p}_i = 1$.
If the initial energy of the particle is higher than $V_B$, then the
work performed is deterministic in the forward and in the backward
process, implying a slope equal to one, as seen in the corresponding region of
Fig.~\ref{fig:actions} b).

\begin{figure}\[
	\includegraphics[scale = 0.6]{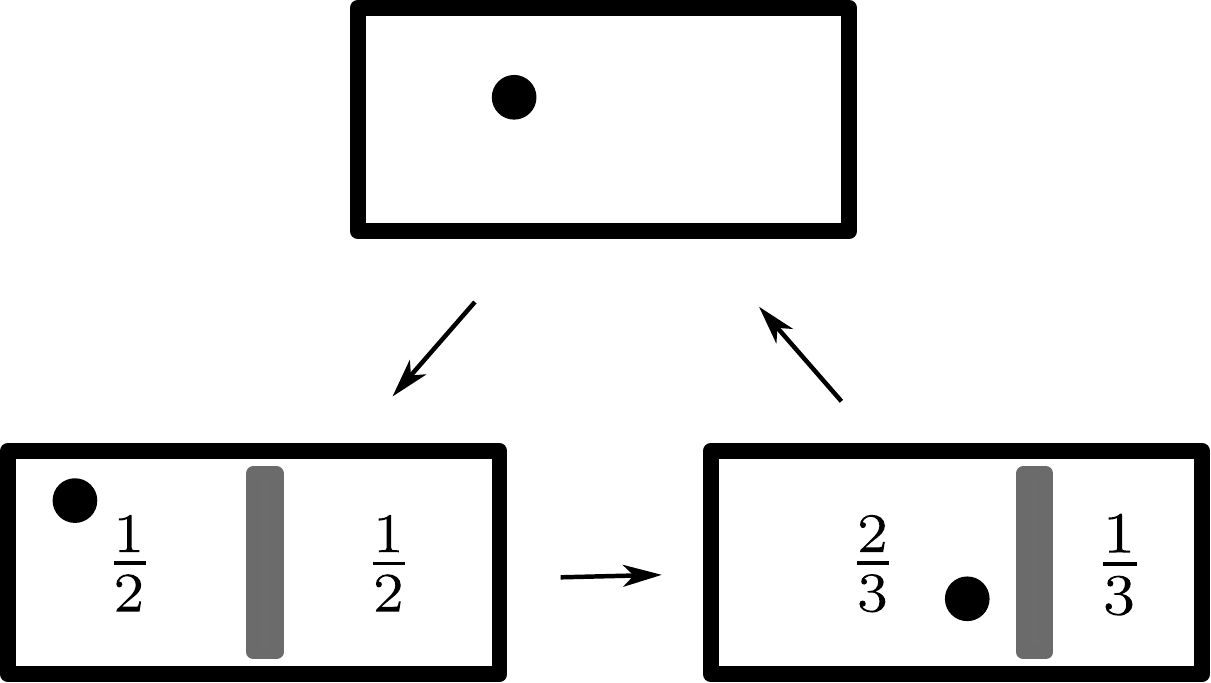}\]
	\caption{Process leading to Fig.~\ref{fig:actions} c). A
classical particle is enclosed in a closed container in weak contact
with a heat bath. At some point, a partition is inserted splitting the
container into two equal parts. The partition is then moved very
slowly to the right until the volume of the left part of the container
is 2/3 of the total volume.}
	\label{fig.ebproc}
\end{figure}

The two previous examples are called {\em microcanonical Szilard engines} because they are able to extract energy from the system in a cyclic protocol, if the process starts with a microcanonical ensemble with energy such that $\phi_{\rm fin}\leq \phi_{\rm init}$.

Our third example, corresponding to 
Fig. \ref{fig:actions} c), is macroscopic: a particle in contact with a thermal bath and following a non-feedback protocol which is a slight modification of the
original Szilard engine.
The system consists of a classical particle in a closed volume, in weak
contact with a large Hamiltonian system (the heat bath).
We perform the following cyclic process: We insert a rigid wall
partitioning the volume accessible to the particle into two
equal parts. Then we
move this wall quasistatically to the right in such a way that, at the end of
the process, the left part occupies $2/3$ of the total volume and the
right part $1/3$ (see Fig.~\ref{fig.ebproc}).
At the end of the process, we remove the wall.

Equation \eqref{slopes} in this case reads
\begin{equation}
	p_i \phi(E) = \tilde{p}_i\phi(E + W_i).
	\label{e.ebsz}
\end{equation}
where $W_i$, $i \in {\rm \{L,R\}}$ is the work performed
on the system, $p_i$ is the probability that the particle is on the side $i$
(left of right) of the wall, and $\tilde{p}_i$ is the probability that
the particle is on the side $i$ of wall in the reverse process.
Hence, in our case, $p_{\rm L} = p_{\rm R} = 1/2$, while
$\tilde{p}_{\rm L} = 2/3$ and $\tilde{p}_{\rm R} = 1 / 3$.

Equation (\ref{e.ebsz}) above can also be interpreted as an adiabatic invariance of
the volume $\phi$  corresponding to each ergodic component  of the phase space.
During the process, the system is ergodic on the part of the
phase space where the particle is on the left side of the moving wall.
Therefore, the volume $\phi_{\lambda,{\rm L}}(E)$ enclosed in the part
of the energy shell where the particle is on the left of the moving
wall, is invariant. Hence, $\phi_{\rm init, L}(E) = \phi_{\rm fin,
L}(E + W_{\rm L})$.
Moreover, $\phi_{\rm init, L}(E) = p_{\rm L} \phi(E)$ and
$\phi_{\rm fin, L}(E+W_{\rm L}) = \tilde{p}_{\rm L}\phi(E +
W_{\rm L})$,
because for one free particle in a closed container of volume $V$, in
weak
contact with a heat bath, we have $\phi(E,V) = V f(E)$,
where $f$ is some function.
Similarly, $\phi_{\rm init, R}(E) = p_{\rm R} \phi(E)$, and $\phi_{\rm
fin, R}(E+W_{\rm R}) = \tilde{p}_{\rm R}\phi(E +
W_{\rm R})$.
In the end:
\begin{equation}
	\phi_{\rm fin} = 
	\left\{
		\begin{array}{ll}
			\phi(E + W_{\rm L}) = \frac{3}{4}\phi(E) = 
			\frac{3}{4}\phi_{\rm init} &\quad
			\textrm{if the particle is in the left side}\\
			\\
			\phi(E + W_{\rm R}) = \frac{3}{2} \phi(E) = 
			\frac{3}{2}\phi_{\rm init} &\quad \textrm{if the
			particle is in the right side}
		\end{array}
		\right.
	\label{e.phif}
\end{equation}
as depicted on Fig.~\ref{fig:actions} c).

The transformation of the phase space volume $\phi(E)$ determines the work performed along the process.
In the limit where the heat bath part is very large,
i.e.~$W_{ {\rm L},{\rm R}} \ll E$, Eq.~\eqref{e.phif} yields
\begin{equation}
	W_{\rm L} = kT \ln \frac{3}{4} < 0,
	\label{e.wl}
\end{equation}
and
\begin{equation}
	W_{\rm R} = kT \ln \frac{3}{2} > 0,
	\label{e.wr}
\end{equation}
which is the result given by classical isothermal thermodynamics. The average work  is $\langle W_i\rangle= kT\ln (3/\sqrt{8})$, which is strictly positive.

This argument can be extended to any process where the probabilities $p_i$ and $\tilde p_i$ do not depend on the energy $E$ and there are no swaps of phase space regions. In that case, Eq.~\eqref{slopes} implies that the maps ${\cal U}_i(\phi)$ are linear, that is,
\begin{equation}\label{slopes2}
\frac{\phi_{\rm fin}}{\phi_{\rm init}}=\frac{\phi_{\lambda_\tau}(E+W_i(E))}{\phi_{\lambda_0}(E)}
=\frac{p_i}{\tilde p_i}
\end{equation}
For a large system, if $W_i(E)\ll E$, we can expand the logarithm of the enclosed volume obtaining
\begin{equation}\label{work0}
\ln \phi_{\lambda_\tau}(E)-\ln \phi_{\lambda_0}(E)+\frac{W_i(E)}{kT}=\ln \frac{p_i}{\tilde p_i}
\end{equation}
which yields, for a cyclic process,  $W_i=kT\ln p_i/\tilde p_i$.

To interpret this equation for non cyclic processes, let us calculate the total enclosed volume assuming weak coupling between the system and the reservoir, i.e., that the Hamiltonian of the total system can be decomposed as ${H}_{\rm tot}(x,x_{\rm B};\lambda)=H_{\rm B}(x_{\rm B})+H_{\rm sys}(x;\lambda)$, where $x$ and $x_{\rm B}$ denote micro states of the system and the reservoir, respectively. The enclosed volume is
\begin{eqnarray}
\phi_\lambda(E)&=&\int dx_{\rm B}\int dx\, {\rm \Theta}\left[E-H_{\rm B}(x_{\rm B})-H_{\rm sys}(x;\lambda)\right] \nonumber \\
&=& \int dx\,\phi_{\rm B}( E-H_{\rm sys}(x;\lambda))\nonumber \\
&=& \phi_{\rm B}(E)\int dx\,e^{-\beta H_{\rm sys}(x;\lambda)}=\phi_{\rm B}(E)Z_\lambda
\end{eqnarray}
where $Z_\lambda$ is the partition function of the system corresponding to Hamiltonian $H_{\rm sys}(x;\lambda)$. The {\em bulk entropy} can be written as
\begin{equation}
S_{\rm bulk}(E)\equiv k\ln \phi_{\lambda}(E)=k\ln \phi_{\rm B}(E)-\frac{F(\lambda)}{T}
\end{equation}
where $F(\lambda)=-kT\ln Z_\lambda$ is the free energy of the system at temperature $T$. Equation \eqref{work0} for the work then reads
\begin{equation}\label{wifin}
W_i(E) - \left[ F(\lambda_\tau)- F(\lambda_0)\right]=kT\ln \frac{p_i}{\tilde p_i}.
\end{equation}
This is a well known relationship in stochastic thermodynamics, first introduced for trajectory bundles in \cite{Kawai:2007kc}. Notice that here we have derived this equation for single trajectories: Eq.~\eqref{slopes2} is valid in general for a trajectory starting with energy $E$ and passing through the ergodic component $i$, whereas \eqref{wifin} can be seen as a way of rewriting \eqref{slopes2} for $W_i\ll E$ and using the standard definitions of free energy and temperature for big systems. Finally, since both $p_i$ and $\tilde p_i$ are normalized, averaging \eqref{slopes2} over $p_i$ one can check that $\langle W_i\rangle\geq  F(\lambda_\tau)- F(\lambda_0)$, i.e., the only way of extracting work is by decreasing the free energy of the system, in conformance with the second law.

Therefore, it is not possible to extract work in a cycle if $p_i$ and $\tilde p_i$ are constant and ${\cal U}_i$ is linear. We need either a swap of phase space regions, as in Fig.~\ref{fig:actions} a), or energy-dependent probabilities $p_i(E)$ and $\tilde p_i(E)$, as in Fig.~\ref{fig:actions} b). It is an open problem to reproduce this type of transformation in a macroscopic system, although it is not hard to find examples where $p_i(E)$  depends on the energy. This occurs, for instance, when a critical point depends on temperature, which is the case for almost every phase transition in nature. Remarkably, exceptions are the symmetry breaking transitions induced by pistons or infinite barriers confining particles, which are the preferred ones in the Szilard engine literature.

\section{Conclusions: from ensembles to single systems}\label{sec:conc}

Along the paper we have analyzed trajectories under feedback and non
feedback quasistatic cyclic protocols breaking ergodicity.  We have
seen that it is possible to break the adiabatic invariance of the
volume $\phi(E)$ enclosed by an energy shell $\gamma_E$, and to reduce
$\phi(E)$ in a quasistatic process. This can be done either by a feedback loop as described in section \ref{sec:liouville} or through a
clever protocol as described in section \ref{sec:helm}. In the latter
case, it is even possible to systematically reduce the energy of a
thermally isolated  Hamiltonian system during a cyclic process, see
Figs.~\ref{fig:actions} a) and b).  The examples known so far have
only a few degrees of freedom, but this idea could in principle be
implemented in a macroscopic system. Finding such a system or, alternatively, proving its impossibility, would help to get a better understanding of the foundations of
statistical mechanics and of the second law.

The loss of ergodicity poses another fundamental question regarding the
definition of entropy. If two meso- or macroscopic phases coexist,
then the behavior of a single system significantly differs from the
behavior of the
ensemble. A single system chooses between one of the phases, whereas
the ensemble can cover the two phases simultaneously.
Moreover, the
ensemble is not necessarily in global equilibrium, since the populations in
each phase depend on the whole process and not only on the value of the
external parameters at a given time \cite{Horowitz2013b}.
For a single
system in a microstate that lies in one of the disconnected energy
layers, the set of available microstates is the corresponding
sublayer and not the entire energy layer.
The volume of this set can suddenly decrease due to a symmetry
breaking, like the one resulting from the insertion of the piston in
the original Szilard engine or the raising of the barrier in the
microcanonical Szilard models described above.
This amounts to a decrease of the
effective entropy of the system which is not accompanied by extra work
or dissipation \cite{Roldan2014,parrondo2001szilard}.
The decrease is counterbalanced if we want to drive the
system back to its initial state, as it happens in the Szilard engine
by virtue of the Landauer's principle. However, if the cycle is not
completed, the reduction of the phase space volume of the available set of
microstates remains. We believe that this reduction should be taken
into account when the second law is applied to single (isolated)
systems.

One extreme example is the whole universe. It is customary to
extrapolate the second law and talk about the increase of entropy in
the universe as a fundamental law. Nevertheless, the universe has
undergone massive symmetry breaking transitions, like nucleogenesis.
The corresponding reduction of the available set of microstates has
not been considered up to our knowledge.
Another example is the appearance of slow variables like the position
of a crystal, which is also the result of a symmetry breaking.

As we have seen along the paper, the loss of ergodicity in Hamiltonian
systems introduces novel aspects that affect the significance and
scope of the second law. We have discussed here some of them:
the evolution in phase space of thermal and microcanonical Szilard
engines,
the
possibility of extracting work from systems in a microcanonical state,
and
the problem of defining entropy for single systems instead of
ensembles.
All this indicates that symmetry breaking transitions and the loss of
ergodicity play an important role in the foundations of statistical
mechanics that deserves further and deeper research.

\section{Aknowledgements}

This work has been financially supported by Grant ENFASIS (FIS2011-22644, Spanish Government).

\bibliography{NatPhysReview}

\begin{thebibliography}{17}

\bibitem{Leff}
H.S. Leff, A.F. Rex, eds., \emph{Maxwell's Demon: Entropy, Information,
  Computing} (Princeton University Press, New Jersey, 1990)

\bibitem{Landauer1961}
R.~Landauer, in \emph{Maxwell's Demon: Entropy, Information, Computing}
  (Princeton University Press, New Jersey, 1990)

\bibitem{Bennett1982b}
C.~Bennett, Int. J. Theor. Phys. \textbf{21}, 905 (1982)

\bibitem{Sagawa2013inbook}
T.~Sagawa, M.~Ueda, \emph{Information Thermodynamics: Maxwell's Demon in
  Nonequilibrium Dynamics} (Wiley-VCH, 2013), pp. 181--211, ISBN 9783527658701

\bibitem{Sekimoto}
K.~Sekimoto, \emph{Stochastic Energetics}, Vol. 799 of \emph{Lect. Notes Phys.}
  (Springer, Berlin Heidelberg, 2010)

\bibitem{gaveau_relative_2014}
B.~Gaveau, L.~Granger, M.~Moreau, L.S. Schulman, Entropy \textbf{16}, 3173
  (2014)

\bibitem{parrondo_thermodynamics_2015}
J.M.R. Parrondo, J.M. Horowitz, T.~Sagawa, Nat Phys \textbf{11}, 131 (2015)

\bibitem{Berut2011}
A.~Berut, A.~Arakelyan, A.~Petrosyan, S.~Ciliberto, R.~Dillenschneider,
  E.~Lutz, Nature \textbf{483}, 187 (2011)

\bibitem{Toyabe2010}
S.~Toyabe, T.~Sagawa, M.~Ueda, E.~Muneyuki, M.~Sano, Nature Phys. \textbf{6},
  988 (2010)

\bibitem{Koski2014b}
J.~Koski, V.~Maisi, T.~Sagawa, J.P. Pekola, Phys. Rev. Lett. \textbf{113},
  030601 (2014)

\bibitem{Roldan2014}
E.~Rold\'an, I.A. Mart\'{\i}nez, J.M.R. Parrondo, D.~Petrov, Nature Phys.
  \textbf{10}, 457 (2014)

\bibitem{Marathe:2010fx}
R.~Marathe, J.M.R. Parrondo, Phys. Rev. Lett. \textbf{104}, 245704 (2010)

\bibitem{Vaikuntanathan:2011en}
S.~Vaikuntanathan, C.~Jarzynski, Phys. Rev. E \textbf{83}, 061120 (2011)

\bibitem{parrondo2001szilard}
J.M.R. Parrondo, Chaos \textbf{11}, 725 (2001)

\bibitem{Kawai:2007kc}
R.~Kawai, J.M.R. Parrondo, C.V. Den~Broeck, Phys. Rev. Lett. \textbf{98}, 80602
  (2007)

\bibitem{Sagawa2009}
T.~Sagawa, M.~Ueda, Phys. Rev. Lett. \textbf{102}, 250602 (2009)

\bibitem{Horowitz2013b}
J.M. Horowitz, J.M.R. Parrondo, Acta. Phys. Pol. B \textbf{44}, 803 (2013)

\end{thebibliography}

\end{document}